\documentclass[aps,11pt]{revtex4} 
\usepackage[dvips]{epsfig} 
\usepackage[english]{babel} 
\usepackage{amsmath} 
\usepackage{amssymb} 

\def\be{\begin{equation}} 
\def\ee{\end{equation}} 
\def\disp{\displaystyle}

\parskip=\medskipamount

\begin{document} 

\title{Topological percolation on a square lattice} 
\author{S.K. Nechaev$^{1,2}$, O.A. Vasilyev$^{3}$} 
\affiliation{$^1$LPTMS, Universit\'e Paris Sud, 91405 Orsay Cedex, France \\ 
$^2$ LIFR-MIIP, Independent University, Moscow, Russia \\ 
$^3$L.D. Landau Institute for Theoretical physics, 117334, Moscow, Russia}

\begin{abstract} 

We investigate the formation of an infinite cluster of entangled threads in a 
(2+1)--dimensional system. We demonstrate that topological percolation belongs 
to the universality class of the standard 2D bond percolation. We compute the 
topological percolation threshold and the critical exponents of topological 
phase transition. Our numerical check confirms well obtained analytical results. 

\end{abstract} 

\maketitle 

\section{Introduction} 

In this paper we would like to link together two different subjects, percolation 
and topology, which since now were considered separately in physical literature. 
The problem discussed below deals with entanglement of threads: we apply to the 
threads a random set of operators of "windings" and check whether two initial 
sets of threads become mutually entangled. The transition between entangled and 
nonentangled states may be considered as a "topological phase transition". We 
investigate the critical properties of such transition and show that it can be 
mapped onto the percolation problem. We also numerically investigate the 
critical exponent of the topological percolation and demonstrate that it is 
equal to the one of the two dimensional percolation.

Let us remind the general formulation of the bond percolation problem. Consider 
a planar (for example, square) lattice of size $L\times L$ ($L\to\infty$) with 
elementary cell of size $1\times 1$. Each bond on the lattice is occupied with 
the probability $p$ independently on all other bonds (i.e. with the probability 
$1-p$ the bond is left empty). There exists a critical value $p=p_{c}$ at which 
the infinitely large connected cluster of occupied bonds is formed. In other 
words, above $p_{c}$ at least two bonds belonging to opposite sides of the 
lattice are connected by a path of occupied bonds. It is known that for the two 
dimensional bond percolation $p_{c}=\frac{1}{2}$ \cite{SA}. The 2D percolation 
has been extensively studied during last decade. The scaling behavior of 
crossing probability (the probability to percolate in, say, horizontal direction 
as a function of $p$) in the vicinity of $p_{c}$ is defined by the correlation 
length exponent \cite{HL1,HL2,HL3}. A deep insight of scaling at the critical 
point has been achieved using the conformal approaches 
\cite{Lang1,Lang2,Lang3,Cardy0,Cardy1,Watts}.

\begin{figure}[ht] 
\epsfig{file=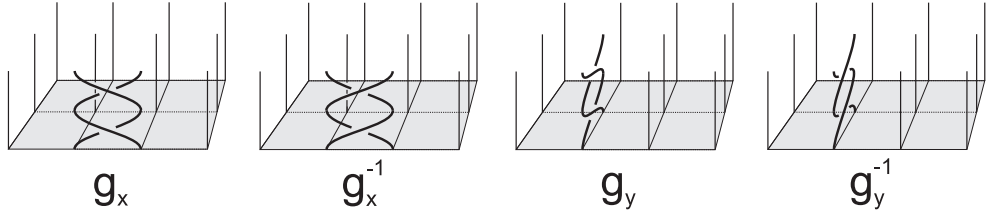,width=13cm} 
\caption{Generators of elementary entanglements of neighboring threads.}
\label{fig:1} 
\end{figure}

The problem discussed below deals with entanglement of threads of finite length 
growing from a two-dimensional substrate. Apparently for the first time the 
continuous version of the same problem has been treated in \cite{Alava1}. Let us 
specify the model under consideration. Consider the two-dimensional square 
lattice of size $L\times L$, serving as the "substrate" from which the threads 
start to grow. Each time moment one pair of nearest neighboring (along $x$-- or 
$y$--axes) lines on the lattice may produce the full-turn entanglement. We 
assign the "generators" $g_x,\, g_x^{-1}$ to the full clockwise and 
counterclockwise turns along $x$--axis, and $g_y,\, g_y^{-1}$ to the full 
clockwise and counterclockwise turns along $y$--axis, and as it is schematically 
shown in the figure \ref{fig:1}.

\begin{figure}[ht] 
\epsfig{file=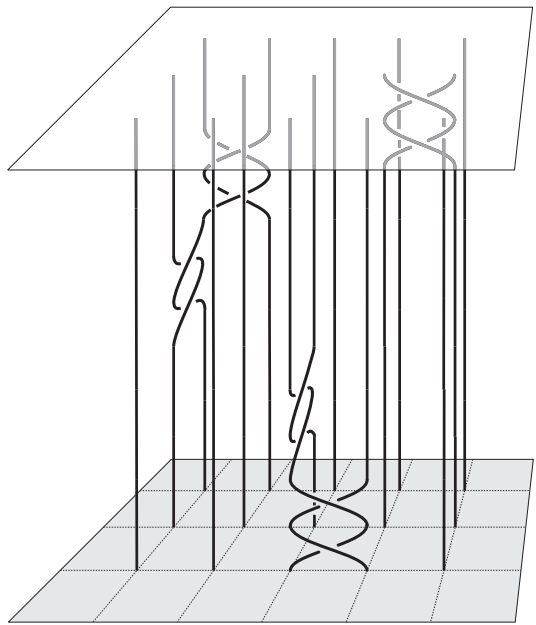,width=7cm} 
\caption{Bunch of entangled threads.}
\label{fig:02} 
\end{figure}

Let us choose randomly any pair of nearest neighboring threads on the lattice 
and entangle them with the probability $q$; with the probability $1-q$ we leave 
the pair of threads unentangled. If we decide to entangle the threads, then we 
do that clockwise with the probability $r^{+}$ and counterclockwise with the 
probability $r^{-}=1-r^{+}$. Later on we shall consider only two extremal cases: 
i) $r^{+}=1$, meaning that all neighboring threads are entangled clockwise 
overall the lattice, and ii) $r^{+}=\frac{1}{2}$, denoting the situation when 
each entanglement may be either clockwise or counterclockwise with equal 
probabilities. Let us stress that for any $r^{\pm}\neq 1$ there is the 
possibility to unwind the neighboring threads, what essentially complicates the 
problem. The general structure of the bunch of entangled directed threads is 
shown in the Fig.\ref{fig:02}.

Now we are in position to formulate the main question of our interest. Suppose 
that each time step we apply randomly one clockwise or counterclockwise winding 
to the top ends of randomly chosen pair of neighboring threads on the lattice of 
size $L\times L$. When time passes, more and more threads become topologically 
connected (entangled). We would like to compute the typical (critical) number of 
applied windings at which the threads on the opposite sides of the substrate 
form the topologically connected cluster, i.e. become entangled. This model sets 
the concept of "topological percolation".

\section{Entanglements and the locally free group} 
The generators $g_x,\, g_x^{-1},\, g_y,\, g_y^{-1}$ of full turns are defined 
for each pair of nearest neighboring threads and, hence, should depend on the 
lattice coordinate (in Fig.\ref{fig:1} the corresponding lattice indices are 
omitted for simplicity): $g_{x}(i,j)$ -- a clockwise ($+$)--generator and $g^{-
1}_{x}(i,j)$ -- a counterclockwise ($-$)--generator of threads $(i,j)$ and 
$(i+1,j)$ along $x$--axis; $g_{y}(i,j)$ -- a clockwise ($+$)--generator and 
$g^{-1}_{y}(i,j)$ -- a counterclockwise ($-$)--generator of threads $(i,j)$ and 
$(i+1,j)$ along $y$--axis. The pair of indices $(i,j)$ is attributed to the 
$(x,y)$ coordinates of a left or a bottom threads in a pair; $i$ and $j$ run 
along $x$ and $y$ axes correspondingly. 

The whole set of generators $\{g_{x}(i,j),\, g_{x}^{-1}(i,j),\, g_{y}(i,j),\, 
g_{y}^{-1}(i,j)\}$ (for all $1\le \{i,j\}\le N-1$) sets the surface locally free 
group ${\cal LF}_{2D}$ \cite{paris,ne_bi}. It is known \cite{ve_ne_bi} that each 
generator $g$ can be represented in a form $g=\sigma^2$ (we have omitted indices 
for brevity), where $\sigma$ is the braid group generator. The generators $g$ 
obey the following commutation relations:
\be 
\left\{\begin{array}{llll} g^{s}_{x}(i,j) g^{s}_{x}(i',j') & = & 
g^{s}_{x}(i',j') g^{s}_{x}(i,j) & \mbox{if $|i-i'|>1$ or $|j-j'|>0$} \medskip \\ 
g^{s}_{y}(i,j) g^{s}_{y}(i',j') & = & g^{s}_{y}(i',j') g^{s}_{y}(i,j) & \mbox{if 
$|i-i'|>0$ or $|j-j'|>1$} \medskip \\ g^{s}_{x}(i,j) g^{ \pm s}_{y}(i',j') & = & 
g^{\pm s}_{y}(i',j') g_{x}(i,j) & \mbox{if $i-i'\ne 0,1$ or $j'-j \ne 0,1$} 
\medskip \\ g^{s}_{x}(i,j) g^{-s}_{x}(i,j) & = & g^{s}_{y}(i,j) g^{-
s}_{y}(i,j)=1 & 
\end{array} \right. 
\label{eq:comm} 
\ee 
where we have attributed $s="+"$ for $g$ and $s="-"$ for $g^{-1}$.

We can reformulate our geometrical problem of entanglement of threads in terms 
of random walk on the group ${\cal LF}_{2D}$. Consider the square lattice of 
size $L\times L$. There are $N=2L^{2}$ threads passing through the vertices of 
the lattice along $z$--direction normal to the plane $(xy)$. Here and later we 
assume the square geometry, but we can consider any other lattices in the same 
way. We generate the random word in terms of "letters"--generators of the group 
${\cal LF}_{2D}$. Namely, we randomly apply one after another generators of 
${\cal LF}_{2D}$ (with appropriate probabilities) to the open ends of threads, 
respecting the commutation relations (\ref{eq:comm}). Let us repeat that the 
generators acting along $x$-- and $y$--axes have equal probabilities. The ($+$) 
and ($-$)--generators have correspondingly the probabilities $r^{+}$ and $r^{-
}=1-r^{+}$.

When a random sequence of $M$ "letters" is generated, we check whether there are 
at least two threads on opposite sides of the lattice which are topologically 
connected to each other (i.e. whether they belong to the same cluster of 
entangled lines). We are interested in statistics of such cross--lattice 
entanglements averaged over different realizations of random sequences of 
"letters". As one sees later, this problem has straightforward relation to the 
two-dimensional bond percolation on the same lattice.

\section{Geometry of entanglements and bond percolation} 

\begin{figure}[ht] 
\epsfig{file=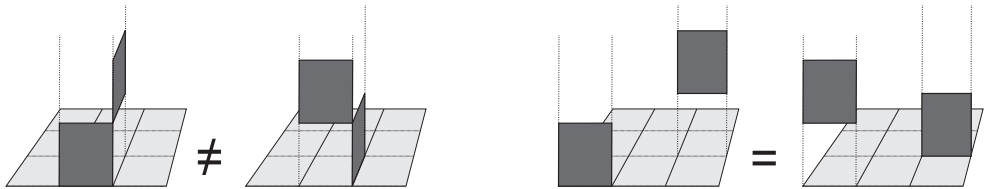,width=13cm} 
\caption{Visualization of some commutation relations in the group ${\cal 
LF}_{2D}$.} \label{fig:2} 
\end{figure}

Let us describe a simple geometric model which visualizes the commutation 
relations (\ref{eq:comm}) of the surface locally free group, making them very 
transparent. We associate the "white" cells to the generators $g_{x}(i,j)$ and 
$g_{y}(i,j)$ and "black" cells to the inverse generators $g_{x}^{-1}(i,j)$ and 
$g_{y}^{-1}(i,j)$. We drop cells randomly, one cell per each time step. The 
commutation relations (\ref{eq:comm}) specify the geometry of the heap 
constituted by falling cells. For example, if the falling cell and one of the 
previous cells have only one common edge, the upper cell remains on the lower 
one; if however the cell falls strictly on the cell with opposite color, these 
two cells annihilate. The figure (\ref{fig:2}) shows the particular example of 
two pairs of non-commuting (left) and commuting (right) generators. 

\begin{figure}[ht] 
\epsfig{file=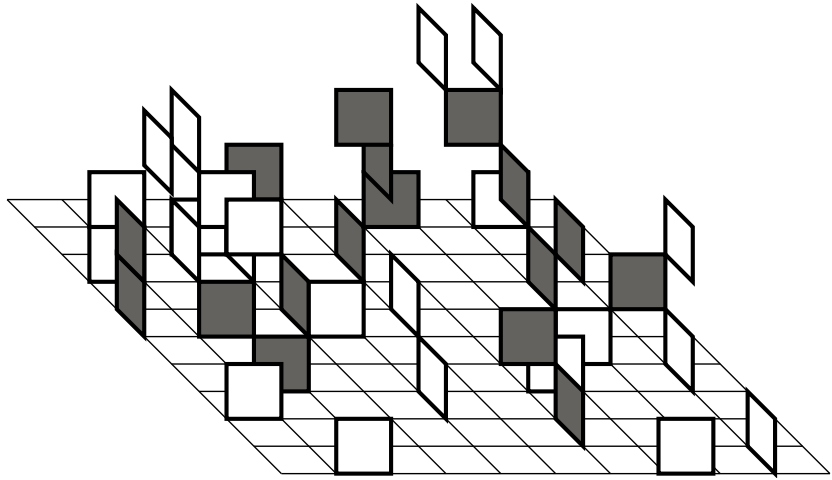,width=11cm} 
\caption{Typical heap of colored cells representing clusters of entangled 
threads.} 
\label{fig:3} 
\end{figure}

The general view of growing heap of white--black cells is shown in 
Fig.\ref{fig:3}. This picture demonstrates the typical configuration of cells 
for the system size $L=10$ and $r^{+}=0.5$. 

Remind that any white (black) cell corresponds to the full clockwise 
(counterclockwise) turn of neighboring vertical threads. The empty horizontal 
intervals between cells mean that the threads in these intervals are not 
entangled.

Consider now the projection of the heap of cells onto the horizontal $xy$--plane 
(the base surface). In this projection the single thread is represented by a dot 
(a lattice site). Each falling cell corresponds to the bond between two 
neighboring sites. If at least one cell in a projection (independent on the cell 
color) covers some bond, we close this bond. Otherwise we leave this bond open. 

It is not difficult to establish the bijection between entanglements and the 
standard percolation. Two distant threads are linked together (entangled) by 
some sequence of full turns of nearest neighboring threads if and only if the 
lattice sites corresponding to these threads belong to the same connected 
cluster of closed bonds. Otherwise these distant threads are not entangled. 
Thus, if a cluster of connected sites touches the opposite sides of the lattice, 
then we can always find two threads on opposite sites of the lattice entangled 
to each other.

There are two formulations of the classical theory of bond percolation: {\it 
canonical} and {\it grand canonical}. In the canonical approach the averaging is 
performed over configurations with preserved number of closed bonds $S$ on the 
lattice and, hence, the normalized concentration of closed bonds $s=\frac{S}{N}$ 
is also fixed. In the grand canonical approach only the probability $p$ of a 
bond to be closed is fixed, while the concentration $s$ fluctuates near the 
average value $\left<s\right>=p$. These two formulations are linked together in 
a simple way. If we know, for example, the crossing probability in the 
horizontal direction, $\pi_{h}(S)$, for the canonical distribution, we can 
obtain the same quantity in the grand canonical formalism by averaging 
$\pi_{h}(S)$ over the Binomial distribution 
\be 
\pi_{h}(p,L)= \sum \limits_{S=0}^{N} P_{b}(S;p,N) \pi_{h}(S,L) 
\label{eq:con} 
\ee 
where $P_{b}(S;p,N)$ is the probability to have exactly $S$ closed bonds on the 
lattice of size $L$ with $N=2L^{2}$ bonds for fixed value of the bond closure 
probability $p$: \be P_{b}(S;p,N)=\frac{N!}{S!(N-S)!}\,p^{S}(1-p)^{N-S} 
\label{eq:bin} 
\ee 
Here and below we assume that a function of the argument $S$ (or of $s$) 
corresponds to the canonical distribution, while a function of the argument $p$ 
corresponds to the grand canonical distribution. The capital letters $S$ and $M$ 
denote the number of closed bonds and number of dropped cells for the entire 
lattice, while the letters $s=\frac{S}{N}$ and $m=\frac{M}{N}$ denote the same 
quantities normalized per number of bonds.

In Refs. \cite{NZ1,NZ2,NZ3} the relation (\ref{eq:con}) has been used for very 
efficient numerical simulation of the grand canonical crossing probability by 
using the canonical one, which is more fast and simple in practical 
applications.

Let us note that eq.(\ref{eq:con}) sets the general relation between two 
different ensembles (grand canonical and canonical) and we shall exploit this 
relation for our goals to establish connection between the standard grand 
canonical formulation of crossing probability and our canonical formulation of 
percolation produced by growing heap. Namely, in our model for each lattice size 
$L$ and specified value $r^{+}$ we randomly add each time moment the new cell to 
the heap and stop at the time moment when $M$ cells are added. After that we 
check the horizontal spanning property of obtained configuration of closed bonds 
by using the Hoshen-Kopelman algorithm \cite{HK}. Then we perform averaging over 
$10^{4}$ different configurations and obtain the topological crossing 
probability $\pi_{t}(M;r^{+},L)$. The only difference from the usual crossing 
probability $\pi_{h}(p,L)$ is that we generate each configuration of closed 
bonds by adding cells to the heap instead of closing each bond independently. 
The crucial difference appears for $r^{+}\ne 0;\,1$, i.e. when annihilation is 
allowed. Let us remind that for topological percolation each closed bond is the 
projection of one (or several) cells from the heap to the base surface. 

\begin{figure}[ht] 
\begin{center} 
\epsfig{file=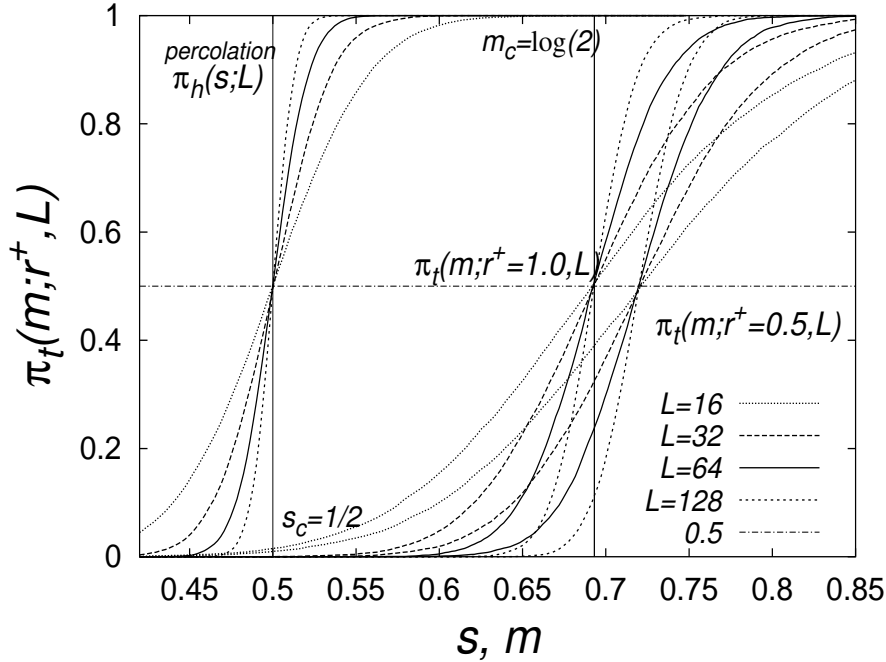,width=12cm,height=9cm} 
\caption{The topological crossing probability $\pi_{t}(m;r^{+},L)$ as a function 
of $m=M/N$ for $r^{+}=1;\,0.5$ and crossing probability $\pi_{h}(s;L)$ of bond 
percolation, canonical case.} 
\label{fig:ph} 
\end{center} 
\end{figure}

In the Fig.\ref{fig:ph} we have plotted the topological crossing probability 
$\pi_{t}(m;r^{+},L)$ as a function of $m=\frac{M}{N}$ for two cases 
$r^{+}=1;\,0.5$. In the same figure the crossing probability $\pi_{h}(s;L)$ of 
the standard bond percolation is plotted for comparison as a function of $s$. 
The graph for the usual 
percolation $\pi_{h}(s;L)$ crosses the horizontal line $0.5$ at the critical 
point $s_{\rm c}=\frac{1}{2}$. The slop of $\pi_{h}(s;L)$ is maximal at the 
critical point. The same construction can be used for the topological crossing 
probability. The graph $\pi_{t}(m;r^{+},L)$ for $r=1$ crosses the horizontal 
line $s_{\rm c}\frac{1}{2}$ at the point with the maximal slop, and the graphs 
for different lattice sizes $L$ cross each other at that point. As we see later, 
this is the critical point of topological percolation for $r^{+}=1$. The the 
similar behavior (however with different critical point) is valid for $r=0.5$.

We can express the topological probability $\pi_{t}(M;r^{+}=1,L)$ via the 
crossing probability $\pi_{h}(M;L)$ in the way similar to eq. (\ref{eq:con}). 
Thus our next goal is to find the probability distribution $P_{t}(S;M,N)$ of 
having $S$ closed bonds on the $N$--bond lattice if we have added $M$ cells to a 
heap. This probability distribution $P_{t}(S;M,N)$ plays the same role for the 
topological percolation as the binomial distribution $P_{b}(S;p,N)$ plays the 
role for the percolation in the "canonical" consideration. To begin with, let us 
consider only clockwise local turns ($r^{+}=1$) because in this case there are 
no cancellations of opposite turns, and the problem becomes essentially more 
simple. We denote this case "the irreversible topological percolation". 

Consider some particular configuration on the lattice with $S$ closed bonds and 
apply the mean--field consideration. Adding one extra cell to a heap, we see 
that with probability $s=\frac{S}{N}$ this cell hits closed already bond and 
does not change the number of closed bonds. However with probability $1-s$ the 
new cell hits the open bond and closes it increasing the number of closed bonds 
by one. Therefore, we can write down the master equation for $P_{t}(S;M,N)$ on 
the lattice $L$:
\be 
P_{t}(S+1;M+1,N)=\frac{S+1}{N}P_{t}(S+1;M,N)+ \left(1-\frac{S}{N}\right) 
P_{t}(S;M,N) 
\label{eq:mast} 
\ee 
with the following initial and boundary conditions: 
\be 
\left\{\begin{array}{l} P_{t}(S=0;M=0,N)=1 \\ P_{t}(S=0;M\ge 1,N)=0 \\ 
P_{t}(S=1;M=1,N)=1 \\ P_{t}(S=1;M\neq 1,N)=0 \end{array}\right. 
\label{eq:bound} 
\ee 
It is easy to solve this problem by the method of generating functions. Define 
\be 
Q(t,z)=\sum_{M=0}^{\infty}\sum_{S=0}^{\infty} P(S;M,N) t^S z^M 
\ee 
Eqs. (\ref{eq:mast})--(\ref{eq:bound}) in terms of $Q(t,z)$ read 
\be 
\frac{1}{N} \frac{\partial Q(t,z)}{\partial t}= \frac{1- t z}{t z (1-t)}Q(t,z)- 
\frac{1}{t z (1-t)} \label{eq:gen} 
\ee 
The solution of (\ref{eq:gen}) (properly normalized) is as follows 
\be 
Q(t,z)=(1-t)^{\frac{N(1-z)}{z}}\, {_2F_1}\left(-\frac{N}{z}, 1+N-\frac{N}{z}, 1-
\frac{N}{z}, t\right) 
\label{eq:sol} 
\ee 
where ${_2F_1}(...)$ is the standard hypergeometric function---see, for example 
\cite{abramowitz}.

Now we can straightforwardly compute the mathematical expectation 
$\left<S(M,N)\right>$ and the dispersion $\left<\Delta S^{2}(M,N)\right> 
=\left<S^2(M,N)\right>-\left<S(M,N)\right>^2$: 
\be 
\begin{array}{rll} 
\left<S(M,N)\right> & = & \disp \frac{1}{2\pi i}\oint_C \frac{d z}{z^{M+1}} 
\left[\frac{d Q(t,z)}{d t}\bigg|_{t=1}\right] = 
N\left\{1-\left(\frac{N-1}{N}\right)^M\right\} \medskip \\ 
\left<S(M,N)(S(M,N)-1)\right> & = & \disp \frac{1}{2\pi i}\oint_C \frac{d 
z}{z^{M+1}} 
\left[\frac{d^2 Q(t,z)}{d t^2}\bigg|_{t=1}\right] \medskip \\ 
& = & \disp N(N-1)\left\{1-2\left(\frac{N-1}{N}\right)^M +
\left(\frac{N-2}{N}\right)^M\right\} 
\end{array} 
\label{eq:moments} 
\ee 
where the contour $C$ surrounds the point $z=0$. Therefore 
\be 
\begin{array}{lll} \left<\Delta S^2(M,N)\right> & = & \left<S(M,N)(S(M,N)-
1)\right>+ \left<S(M,N)\right>- \left<S(M,N) \right>^{2} \medskip \\ 
& = & \disp N(N-1)\left(\frac{N-2}{N}\right)^M-N^{2} \left(\frac{N-
1}{N}\right)^{2 M}+ 
N \left(\frac{N-1}{N}\right)^{M}
\end{array} 
\label{eq:ds}
\ee 

Let us evaluate the normalized expectation $\left<s(m,L)\right>= 
\frac{\left<S(M,N)\right>}{N}$ and the dispersion $\left<\Delta 
s^{2}(m,L)\right>= \frac{\left<\Delta S^{2}(M,N)\right>}{N^{2}}$, where 
$N=2L^{2},\;m=\frac{M}{N}$. In the asymptotic regime, where $N\gg 1,\, M\gg 1$ 
and $m=\frac{M}{N}={\rm const}$, we can rewrite (\ref{eq:moments}) for 
normalized quantities in the limit $L \to \infty$ as follows 
\be 
\left<s(m,N)\right>\equiv \frac{\left<S\right>}{N}= 1-
\lim_{N\to\infty}\left[\left(1-\frac{1}{N}\right)^{N}\right]^{m}=1-e^{-m} 
\label{eq:lim}
\ee 
and
\be 
\frac{\left<S^2\right>}{N} = \lim_{N\to \infty} \left[ (N-1)\left(1-
\frac{2}{N}\right)^{N m}- N\left(1-\frac{1}{N}\right)^{2 N m}+ \left(1-
\frac{1}{N}\right)^{N m}\right] = e^{-m}-(1+m)e^{-2m} 
\label{eq:lim2} 
\ee 

On the basis of (\ref{eq:lim2}) we expect for all $r^{+}$ 
\be 
\left<\Delta s^{2}(m;r^{+},L) \right> < \frac{\rm const}{N} =\frac{\rm 
const}{2L^{2}} 
\label{eq:ogr} 
\ee 

Now we are in position to attack the "reversible topological percolation", i.e. 
the case $r^{+}=0.5$ when both clockwise and counterclockwise turns are allowed. 
First of all we estimate the maximal number of annihilated cells. In 
\cite{ne_bi} the average number $\left<H\right>$ of the most top cells (called 
"the roof") of the heap in the stationary regime $M\gg N$ has been analytically 
computed. Only those cells are accessible for cancellations. We reproduce here 
the result of \cite{ne_bi}: 
\be
\left<h(m,N)\right>=\frac{\left<H(M,N)\right>}{N}=\frac{1}{7}
\label{eq:h}
\ee
For convenience we have reproduced in Appendix the derivation of this 
expression.

Following the methods developed in the works \cite{ne_bi,ne_des,des} one can 
also compute the dependence $\left<h(M,N)\right>$ in the intermediate regime 
$M<N$. The function $\left<h(m,N)\right>$ in all regimes is well approximated by 
the curve 
\be
\left<h(m,N)\right>=\frac{1}{7}(1-e^{-7 m})
\label{eq:hm}
\ee 
The analytic derivation of this expression is presented in Appendix. To verify 
(\ref{eq:hm}) we compare in Fig.\ref{fig:hm} the data of numerical simulations 
for the average size of the roof $\left<h(m;r^{+},L)\right>$ for $r^{+}=0.5$, 
$L=128$ with eq.(\ref{eq:hm}). One sees that the function 
$\left<h(M,N)\right>=\frac{1}{7}(1-e^{-7 m})$ perfectly fits the data. 

\begin{figure}[ht] 
\begin{center} 
\epsfig{file=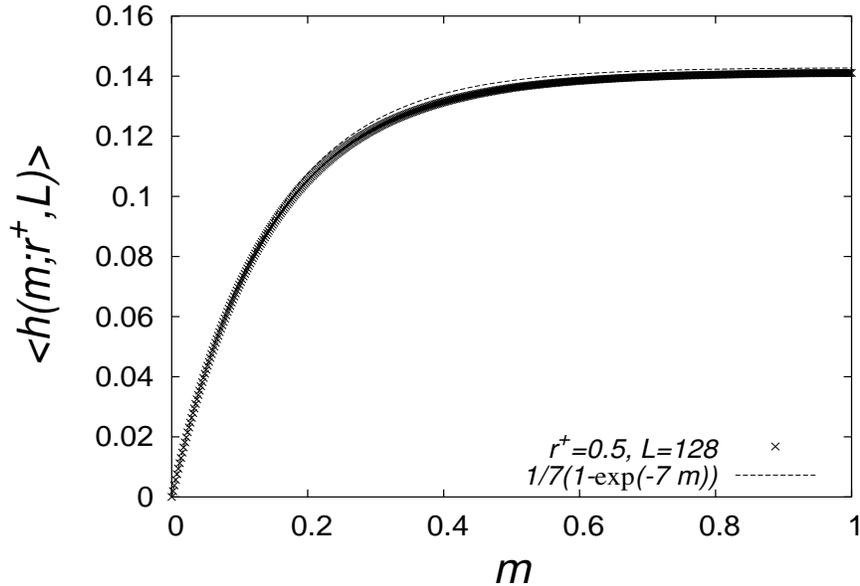,width=12cm,height=8cm} 
\caption{Dependence the size of the roof of the heap $h(m;r^{+},L)$ for 
$r^{+}=0.5$ and $L=128$ in comparison with $\frac{1}{7}(1-e^{-7m})$.} 
\label{fig:hm} 
\end{center} 
\end{figure} 

Let us make one step back and demonstrate that for $r^{+}=1$ we can arrive at 
eq. (\ref{eq:lim}) in a simple way. Suppose that some configuration of cells 
covers the base surface. If we put one more cell, it hits the empty bond and 
increases the number of closed bonds by one with the probability $1-s$, while it 
hits the closed bond and does not change the number of closed bonds with the 
probability $s$. These conditions are summarized in the table below:
\be
\Delta s(m,r^{+}=1)= 
\left\{\begin{array}{rl}
+1 & \mbox{with the probability $1-s$} \medskip \\
0 & \mbox{with the probability $s$} 
\end{array}
\right.
\ee
So, we have 
\be
\left\{\begin{array}{l}
\disp \frac{d \left<s(m,r=1) \right>}{dm}=1-s(m) \medskip \\
s(m=0)=0
\end{array} \right.
\label{eq:dsm}
\ee
The solution to this equation is (\ref{eq:lim}). 

We modify now eq. (\ref{eq:dsm}) to catch the "reversible topological 
percolation" ($r^{+}=0.5$), taking into account the annihilation. We have 
stressed that the annihilation of cells occurs in the roof only. Each bond of 
the roof is either black or white with the probability $\frac{1}{2}$. Hence the 
total probability of annihilation is about 
$\frac{1}{2}\left<h(m;r^{+},L)\right>$. Moreover, the bond in the base surface 
becomes "open" if and only if the cell of the roof is single in the 
corresponding column, i.e. below the given roof's cell there are no other cells 
in the same column. The probability to have empty set of $1-\left<s(m)\right>$ 
columns is given by the solution of eq.(\ref{eq:dsm}): $1-\left<s(m)\right>=e^{-
m}$. Hence, the probability $p_{\rm cond}$ of annihilation of the roof's cell 
under the condition that below this roof's cell there are no other cells in the 
same column (assuming the uniform distribution of the roof's cells and empty 
columns), is given by the product:
\be
p_{\rm cond} = \frac{1}{2} \left<h(m;r^{+}=0.5)\right> \times e^{-m}=
\frac{e^{-m}}{14}(1-e^{-7 m})
\label{eq:cond}
\ee
Therefore 
\be
\Delta s(m,r^{+}=0.5)= 
\left\{\begin{array}{rl}
-1 & \mbox{with the probability $\frac{e^{-m}}{14}(1-e^{-7 m})$} 
\medskip \\
0 & \mbox{with the probability $s - \frac{e^{-m}}{14}(1-e^{-7 m})$} 
\medskip \\
+1 & \mbox{with the probability $1-s$} 
\end{array}
\right.
\ee
Hence, we can write 
\be
\left\{\begin{array}{l}
\disp \frac{d \left<s(m),r^{+}=0.5\right>}{dm}=1-s(m) - \frac{e^{-
m}}{14}\left(1-e^{-7 m}\right) \medskip \\
s(m=0)=0
\end{array}\right.
\label{eq:dsmr}
\ee
The solution to eq. (\ref{eq:dsmr}) reads
\be
\left<s(m;r^{+}=0.5)\right>= 1 - \frac{e^{-8 m}}{98} - \frac{97\, e^{-m}}{98} - 
\frac{m\, e^{-m}}{14}
\label{eq:f}
\ee

\begin{figure}[ht] 
\begin{center} 
\epsfig{file=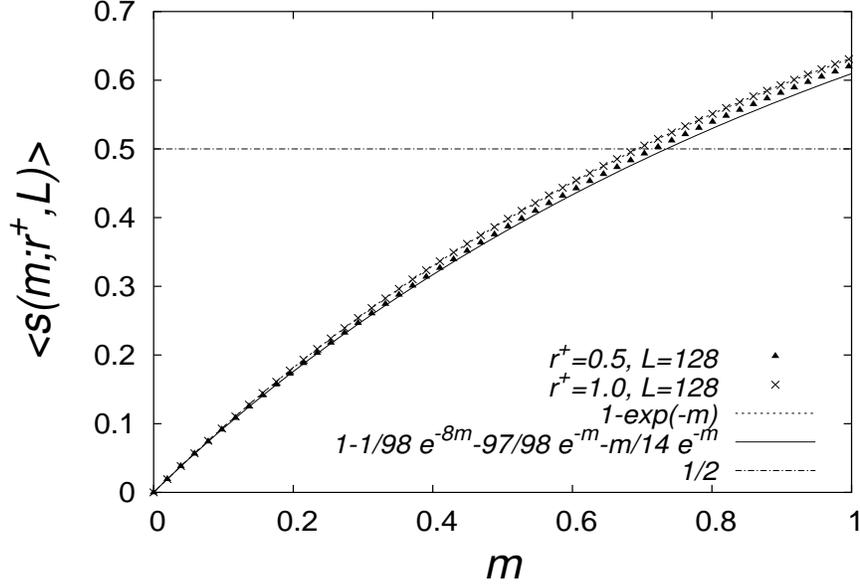,width=12cm,height=8cm} 
\caption{Comparison of $\left<s(m;r^{+},L)\right>$ for $r=1,0.5$ and $L=128$ 
with the approximation functions.}
\label{fig:smr} 
\end{center} 
\end{figure}

In Fig.\ref{fig:smr} we plot the numerical data for $\left <s(m;r^{+},L) 
\right>$ for $r^{+}=1,0.5$ and $L=128$ (crosses). We see, that data for 
$r^{+}=1$ is in the excellent agreement with eq. (\ref{eq:lim}). For comparison, 
in the same figure we plot also the analytic expression (\ref{eq:f}) for the 
"reversible topological percolation". As one sees, this curve fits rather well 
the corresponding numerical simulations. 

Let us introduce the probability distribution for normalized quantities 
$P_{t}(m,s,L)= 2L^{2}\,P_{t}(2L^{2}s,2L^{2}m,2L^{2})$. We can write (at least 
for the critical region) the scaling expression for the crossing probability 
\be 
\pi_{h}(s,L)\simeq f \left((s-s_{c})L^{1/\nu} \right) 
\label{eq:sph}
\ee 
where $\nu=\frac{4}{3}$ is the critical exponent of the correlation length for 
the
percolation. Therefore 
\be 
\left. \frac{\partial^{2} \pi_{h}(s,L)}{\partial s^{2}} \right|_{s\to s_{c}} 
\simeq L^{2/\nu}\, f''((s-s_{c})L^{1/\nu}) 
\label{eq:dsph} 
\ee 
For $S \gg 1$, $N \gg 1$ we have 
\be  
\begin{array}{lll} 
\pi_{t}(m_{c},L) & = & \disp \sum_{S=0}^{N} P_{t}(S;M_{c},N) 
\pi_{h}(S/(2L^{2}),L) 
\simeq \int_{0}^{1} P_{t}(s;m_{c},L) \pi_{h}(s,L) d s \medskip \\ 
& \simeq & \disp \int_{0}^{1} P_{t}(s;m_{c},L) \left(\pi_{h}(s_{c},L) + \left. 
\frac{\partial \pi_{h}(s,L)}{\partial s} \right|_{s=s_{c}}\Delta s + \left. 
\frac{ \partial^{2} \pi_{h}(s,L)}{\partial s^{2}} \right|_{s=s_{c}}\Delta s^{2} 
+
o(\Delta s^{2}) \right) d s \medskip \\ 
& \simeq & \disp \pi_{h}(s_{c},L)+ \left. 
\frac{ \partial^{2} \pi_{h}(s,L)}{\partial s^{2}} \right|_{s=s_{c}} \left<\Delta 
s^{2}(m,L) \right> \simeq \pi_{h}(s,L)+ L^{2/\nu-2}f''(s-s_{c}))+
o(L^{2/\nu-2}) 
\end{array} 
\label{eq:sim}
\ee 
where $\Delta s=s-s_c$, $s_{c}=1-e^{-m_{c}}$ and the function $f(s-s_c)$ is 
regular in the vicinity of the point $s_c$. We use the condition (\ref{eq:ogr}) 
for the dispersion $\left<\Delta s^{2}(m,L)\right>$. Hence for large lattices 
one has
\be 
\lim_{L \to \infty} \pi_{t}(m;\,r=1,\,L)=\pi_{h}\Big(1-e^{-m},\, L\Big) 
\label{eq:res1} 
\ee 

The bond percolation on infinite square lattice occurs at the concentration of 
bonds $s_{c}=p_{c}=\frac{1}{2}$. Hence, using (\ref{eq:lim}), we get the 
critical concentration of local turns (or falling cells) $m_{\rm c}=\frac{M_{\rm 
c}}{N}$ at the percolation threshold in the limit $L\gg 1$ for irreversible 
(($r^+=1$) and reversible ($r^+=0.5$) cases:
\be 
\left\{
\begin{array}{ll}
\disp m_{\rm c}(r^+=1)=\ln 2\simeq 0.693 & \quad \mbox{for $r^+=1$} \medskip \\
\disp m_{\rm c}(r^+=0.5)\approx 0.734  & \quad \mbox{for $r^+=0.5$}
\end{array}
\right.
\label{eq:ent} 
\ee 
These solutions are obtained from eqs. (\ref{eq:lim}) and (\ref{eq:f}) 
correspondingly:
$$
\left\{\begin{array}{l}
s_{\rm c}=\frac{1}{2} = 1-e^{-m_{\rm c}} \medskip \\
s_{\rm c}=\frac{1}{2} = 1 - \frac{e^{-8 m_{\rm c}}}{98} - 
\frac{97\, e^{-m_{\rm c}}}{98} - \frac{m\, e^{-m_{\rm c}}}{14}
\end{array}\right.
$$
The values (\ref{eq:ent}) are in good agreement with the data of numerical 
simulations shown in Fig.\ref{fig:ph}. 

\begin{figure}[ht] 
\begin{center} 
\epsfig{file=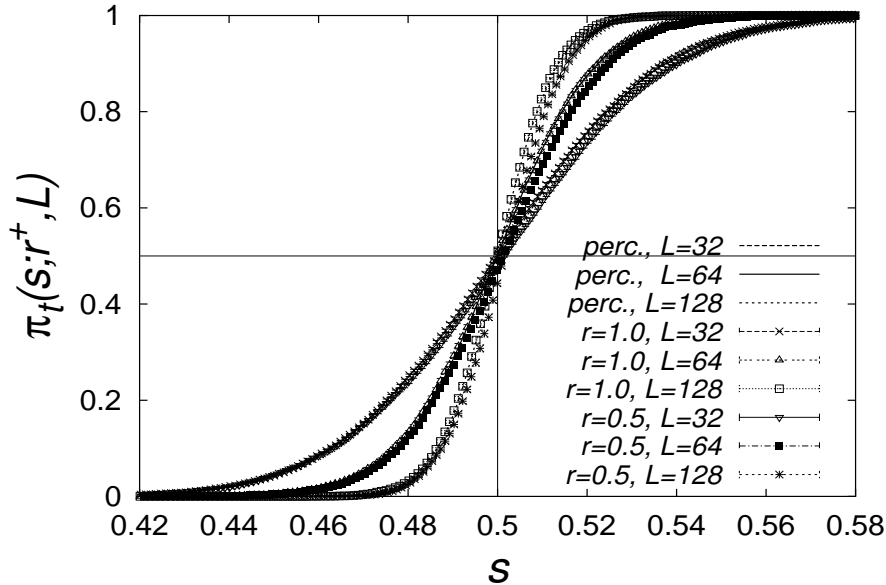,width=12cm,height=8cm} 
\caption{The topological crossing probability $\pi_{t}(s;r^{+},L)$ as a function 
of the average number of the closed bonds $s=1-\exp(-m)$ for $r=0.5,1$ and 
crossing probability $\pi_{h}(s;L)$ of bond percolation, canonical case.} 
\label{fig:phn} 
\end{center} 
\end{figure} 

In the Fig.\ref{fig:phn} we show by symbols the topological crossing probability 
$\pi_{t}$ as a function of a variable $s=1-\exp(-m)$ for $r^{+}=1,0.5$ and 
$L=32,64,128$. In the same figure we plot the crossing probability of bond 
percolation $\pi_{h}(s;L)$ for the canonical case by lines. One sees that the 
symbols for $r^{+}=1$ lie on the appropriate lines. Our investigation allows us 
to conclude that the topological percolation belongs to the universality class 
of the two dimensional percolation. We confirm this statement numerically. 
Namely, we extract the correlation length exponent $\nu=4/3$ from the numerical 
data shown in the Fig.\ref{fig:phn}. The outline of our construction is as 
follows. The eq.(\ref{eq:sph}) allows us to conclude that the derivative of the 
crossing probability at the critical point scales with the lattice size as 
\be  \pi'_{h}(s_{c},L)\simeq f'(0) L^{1/\nu} 
\label{eq:dphs}
\ee 

\begin{figure}[ht] 
\begin{center} 
\epsfig{file=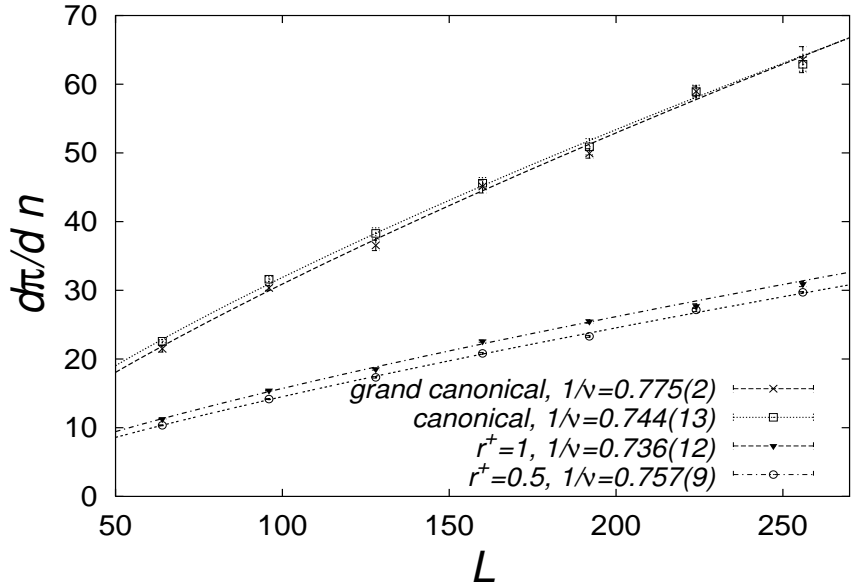,width=12cm,height=8cm} 
\caption{Extraction of the inverse critical exponent of the correlation length 
form the derivative of the crossing probability at the critical point.} 
\label{fig:n2}
 
\end{center} 
\end{figure}

Now we define numerically  the derivative of the crossing probabilities. We 
approximate the data for $\pi_{t}(m;r^{+},L)$ in the vicinity of the critical 
point $0.5-0.05 \le \pi_{t}(m;r^{+},L) \le 0.5+0.05$ by the linear function 
$0.5+a(r,L)(m-m_{c}(r^{+},L))$. We perform this procedure for grand canonical 
and canonical distributions of usual percolation (crossing probabilities 
$\pi_{h}(p,L)$ and $\pi_{h}(s,L)$) as well as for the topological percolation 
$r^{+}=1,0.5$ (crossing probability $\pi_{t}$). Then we plot the derivative of 
the crossing probability at the critical point $\pi'(s_{c},r,L)= a(r,L)$ as a 
function of the lattice size $L$ for grand canonical and canonical distributions 
for the standard percolation, as well as for irreversible ($r^{+}=1$) and 
reversible ($r^{+}=0.5$) topological percolation. The corresponding results are 
shown in Fig.\ref{fig:n2}. 

The data depicted in Fig.\ref{fig:n2} are approximated by the function $a 
L^{1/\nu}$. The result of approximation is plotted in the Fig.\ref{fig:n2} by 
lines. One sees that the numerical values of the inverse correlation length 
length $1/\nu$ coincide with the analytical value $\frac{3}{4}$ within accuracy 
of the approximation. 

\section{Conclusion} 

We have investigated the topological phase transition in the bunch of randomly 
entangled  directed threads. Specifically we pay the most attention to the 
determination of the minimal number $M_{\rm c}$ of local turns necessary to 
produce the fully topologically connected cluster of threads. Namely, below 
$M_{\rm c}$ the opposite side of the square lattice of threads are disjoined, 
while above $M_{\rm c}$ the opposite sides belong to the single cluster of 
connected threads. Two models of topological percolation are considered: 
"irreversible" and "reversible". In the irreversible case ($r^{+}=1$) all local 
turns are only clockwise, while in the reversible case both clockwise and 
counterclockwise local turns are available with equal probabilities 
($r^{+}=0.5$).

We map the problem of topological percolation onto the standard two-dimensional 
percolation and relate the above defined value $M_{\rm c}$ (normalized per the 
number of lattice bonds) to the critical value $p_{\rm c}$ of the percolation 
threshold on the square lattice. The consideration of the reversible topological 
percolation demands  special care. We give the corresponding estimate for the 
critical value $M_{\rm c}$ considering the reversible topological percolation as 
the growth of the random heap of pieces. In particular we estimate the 
probability to "open" the bond of the base surface from the probability of 
cancellation of a piece in a growing heap.

In addition, we find numerically the critical exponent for the topological 
percolation and show that is is equal the correlation length exponent 
$\nu=\frac{4}{3}$ of the standard two-dimensional bond percolation. 

\begin{appendix}
\section{}

The process of growth of a heap (i.e. the random walk on the surface locally 
free group ${\cal LF}_{2D}$) consists in adding step-by-step new "black" or 
"white" blocks to the roof. The dynamics of a heap is controlled by the dynamics 
of a roof. For a particular configuration of a heap we define the number $H$ of 
most top segments in a heap (i.e. the "size of a roof") as well as the number 
"non-roof's" segments, $L_i$, having exactly $i$ neighboring roof's segments---
see \cite{ne_bi}. (Remind that there are $N=2L^{2}$ lattice bonds on the 
lattice). For the values $H,\,L_i$ the following conditions hold:
\be 
\left\{\begin{array}{l}
L_0+L_1+L_2+H=2L^2+2L \\
6 H-8(L+1)\le L_1+2L_2\le 6 H
\end{array}\right.
\label{eq:ap1}
\ee
If $L\gg 1$ one can neglect the boundary conditions and rewrite (\ref{eq:ap1}) 
is simpler form
\be 
\left\{\begin{array}{l}
L_0+L_1+L_2+H = 2L^2 \\
L_1+2L_2 = 6 H
\end{array}\right.
\label{eq:ap2}
\ee

For a given configuration the local dynamics of a size of a roof reads
\be
\left\{\begin{array}{ll}
\Delta H=1 & \mbox {with the probability $\frac{L_0}{2L^2}$} \\
\Delta H=0 & \mbox {with the probability $\frac{L_1+H}{2L^2}$} \\
\Delta H=-1 & \mbox {with the probability $\frac{L_2}{2L^2}$}
\end{array}\right.
\label{eq:ap2aa}
\ee
Equations (\ref{eq:ap2aa}) allow to write the following equation for the 
expectation $\left<H(M)\right>$:
$$
\frac{d\left<H(M)\right>}{dM}=\frac{L_0-L_2}{2L^2}
$$
This equation can be rewritten for the normalized quantities $\left<h(m)\right>= 
\frac{\left<H\right>}{2L^2}$ and $m=\frac{M}{2L^2}$ in the closed form with the 
help of (\ref{eq:ap2}). We get:
\be
\left\{\begin{array}{l}
\disp \frac{d\left<h(m)\right>}{dm}=1-7h \medskip \\
\left<h(m=0)\right>=0
\end{array}\right.
\label{eq:ap2a}
\ee
The solution to (\ref{eq:ap2a}) reads
\be
\left<h(m)\right>=\frac{1}{7}(1-e^{-7m})
\label{eq:ap2b}
\ee
The derivation of (\ref{eq:ap2b}) is the basis for the approximation 
(\ref{eq:hm}). In a stationary case (i.e. for $m\to\infty$) we have 
$\left<h(m)\right>=\frac{1}{7}$. This expression coincides with (\ref{eq:h}).

\end{appendix}

\end{document}